\documentstyle[floats,prd,aps]{revtex}
\begin{document}
\draft
\preprint{UM-P-96/45}
\title{The princess and the pea}
\author{ Neil J. Cornish$^{\dag}$ and Norman E. Frankel$^{\ddag}$} 
\address{${\dag}$DAMTP. University of Cambridge, Silver Street, Cambridge CB3
9EW, UK}
\address{${\ddag}$School of Physics, University of Melbourne, Parkville 3052,
Victoria, Australia}
\twocolumn[
\maketitle
\widetext
\begin{abstract}
Like a fairy-tale princess, trajectories around black holes can be
sensitive to small disturbances. We describe how a small disturbance
can lead to erratic orbits and an increased production of
gravitational waves. 
\end{abstract}
\pacs{04.30.Db, 05.45.+b, 97.60.Lf}
]
\narrowtext

\begin{picture}(0,0)
\put(420,160){{\large UM-P-96/45}}
\end{picture} \vspace*{-0.15 in}

The high degree of symmetry found in the spacetime of an isolated
black hole leads to regular geodesic motion for orbiting bodies.
Working on the assumption that small disturbances generally have small
effects, it is often tacitly assumed that this idealised,
textbook picture carries over to real astrophysical situations. Here we want
to emphasise that the idealised picture is not stable against small
perturbations, as the non-linearity of Einstein's equations tends to
amplify small disturbances.

The observation that small perturbations of
an idealised black hole spacetime can lead to qualitative
changes in the dynamics is not new. It has previously been noted that
a range of perturbations lead to chaotic dynamics. The perturbations
considered include additional mass concentrations
\cite{chandra,conto,moeckel,us,vl}, magnetic fields\cite{kv},
gravitational waves\cite{bc} and spin-orbit coupling\cite{maeda}.
Unlike the Kepler problem of Newtonian mechanics, essentially any
perturbation of an isolated black hole spacetime will lead to
chaotic orbits. This is because even the most pristine
black hole spacetime harbours
the seeds of chaos in the form of isolated unstable orbits. A small
perturbation causes these unstable orbits to break out and infest large
regions of phase space. Note that experience with
Newtonian systems is very misleading. For example, the Kepler problem
has more integrals of motion than are needed for integrability. Keplerian
systems are thus impervious to small perturbations. In contrast,
black hole spacetimes are at the edge of chaos, just waiting for the
proverbial butterfly to flap its wings.

Once it is realised that typical black hole -- satellite systems are
chaotic in the strong field regime, we are forced to consider
the consequences. One of the most immediate consequences is that there
will be no such thing as the ``last stable orbit''\cite{kid}. The
boundary between stable and unstable orbits will be fractal, and there
may be large fractal tendrils of unstable orbits invading what would
have been stable territory in an ideal black hole spacetime. This
feature will be important in determining when a binary system
switches from inspiral to cataclysmic collapse\cite{kid}. Another
important consequence
will be the increased production of gravitational waves due to
the erratic motion of chaotic orbits.

In what follows we will illustrate
both of these effects in a simple model system. While our model does not
describe a real astrophysical situation, it does capture many of the
salient features we expect to find in a relativistic binary system.

For the purpose of illustration we will study the orbits of non-rotating
satellites around an extreme Riessner-Nordstrom black hole. Almost identical
results hold for motion around a Schwarzschild black
hole perturbed by an orbiting third body\cite{moeckel}. Extreme black holes
have the added advantage of allowing an exact generalisation to spacetimes
with $N$ extremal masses\cite{mp}. Rotating black holes bring with them a host
of new instabilities which only amplify the points we wish to make.
Similarly, rotating satellites introduce additional instabilities
through spin-orbit\cite{maeda} and spin-spin couplings.
So while our model is chosen on the grounds of simplicity, it is
likely that more realistic models will be even more sensitive to
small perturbations. Moreover, a central feature of non-linear
dynamics is universality: the details of the dynamics are less
important than the general structure of the phase space
trajectories. For example, a stochastic layer will lead to an increase
in the gravitational wave luminosity regardless of what lead to the
formation of the stochastic layer. Because of this, the effects we
describe for our particular model will enjoy wider applicability.

Our unperturbed spacetime is described by the metric
\begin{equation}
ds^2=-\left(1+{M \over r}\right)^{-2}dt^2+\left(1+{M \over r}\right)^2
\left( dr^2 +r^2 d\Omega^2\right) \, ,
\end{equation}
and electromagnetic potential
\begin{equation}
A_{t}=\left(1+{M \over r}\right)^{-1} \, .
\end{equation}
Here $d\Omega^2=d\theta^2+\sin^2\theta d\phi^2$ is the metric of
a 2-sphere. Into this spacetime we introduce a spinless satellite of mass $\mu$
and consider its orbits. To keep our model simple we will neglect
the spacetime curvature caused by the satellite.
The rotational symmetry allows us to restrict our attention to
orbits in the plane $\theta=\pi/2$. Moreover, invariance under
time translations and spatial rotations leads to conservation
of the satellite's energy $\mu E$ and angular momentum $\mu L$. This allows
the motion to be reduced to the radial equation
\begin{equation}
\left({d r \over d \tau}\right)^2= E^2 - V^2(r) \, ,
\end{equation}
where $\tau$ is the proper time along the trajectory and
\begin{equation}
V(r)=\left(1+{M\over r}\right)^{-1}\left(1+{L^2 \over (r+M)^2}\right)^{1/2}\, .
\end{equation}
We will be concerning ourselves with bound orbits $(E<1)$, but similar
conclusions hold for unbound orbits. In Fig.~1 the effective
potential $V(r)$ is displayed for three values of $L$. The energy $E=0.94$ is
also displayed. We see that some orbits can reach the event horizon at
$r=0$ while others are protected by an angular momentum barrier. For
$E=0.94$ this occurs at the critical angular momentum $L_{c}=2.982M$.
A trajectory with these parameter values can follow an unstable
circular orbit at $r_{c}=(L_{c}^2-2M^2-L_{c}\sqrt{L_{c}^2-8M^2})/2M$. As
emphasised by many authors, it is the existence of such unstable
periodic orbits that make black hole spacetimes sensitive to small
perturbations.

\
\begin{figure}[h]
\vspace{55mm}
\includegraphics{blank.ps}
\includegraphics{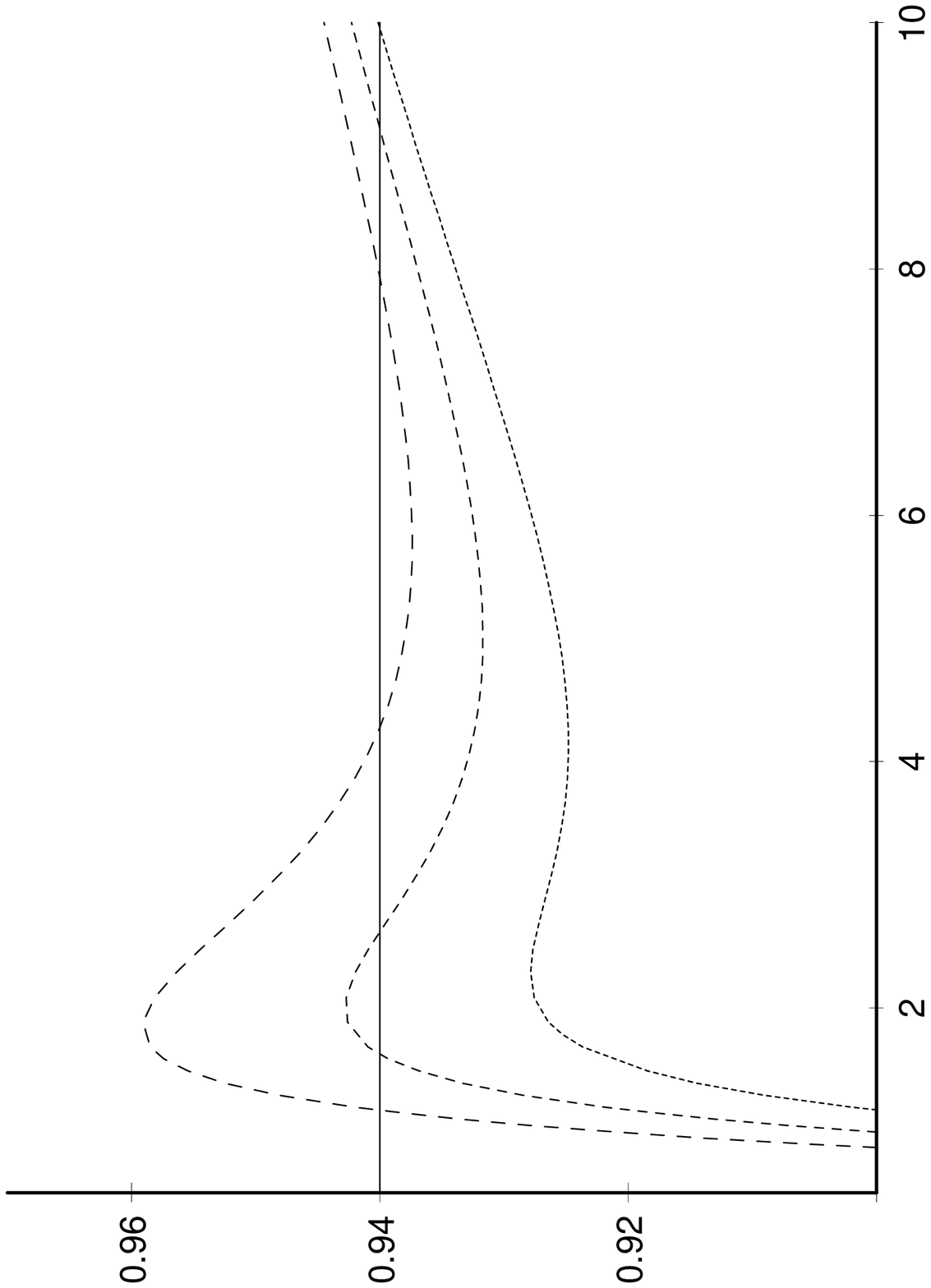}
\vspace{8mm}
\caption{The effective potential $V(r)$ for different values
of the angular momentum. In order of decreasing height, the potentials
correspond to $L=3.1M, \, 3M,$ and $2.9M$ respectively. The solid
line is the fixed energy $E=0.94$.}
\end{figure} 

\begin{picture}(0,0)
\put(5,245){$V$}
\put(120,80){$r/M$}
\end{picture}

Previous studies have used local methods to assess
the onset of instability. Here we apply a more revealing global approach
based on stability basins. Such global techniques are increasingly being
used in engineering applications in order to survey the full range of
stable and unstable configurations of circuits, bridges and boats\cite{eng}.
When systems are chaotic, stable
regions of phase space can be invaded by chaotic tendrils. These
tendrils exhibit a complicated fractal structure.

To illustrate how small perturbations can lead to important changes in
the dynamics we introduce a small ``pea'' with mass $m$ and charge
$q=m$ at a distance $r=R$ from the black hole. The new metric is
obtained by replacing $1+M/r$ by
\begin{equation}
U=1+{M \over r}+{m \over \sqrt{R^2-2Rr\cos\phi+r^2} }\, ,
\end{equation}
leading to
\begin{equation}
ds^2=U^{2}\left(E^2 d\tau^2+ dr^2 +r^2d\theta^2+\sin^2\theta d\phi^2\right)\, .
\end{equation}
The new metric is an exact solution to the Einstein-Maxwell field
equations belonging to the Majumdar-Papapetrou\cite{mp} family of spacetimes.
The leading corrections to the extreme Riessner-Nordstrom metric are given by
\begin{equation}
\Delta U = {m \over R}\left(1+{r \over R}\cos\phi+{r^2 \over 2R^2}
\left[3\cos^2\phi-1\right]+\dots \right) .
\end{equation}
The dipole term, $\cos\phi$, can be removed by a coordinate
transformation, while the quadrupole term, $(3\cos^2\phi -1)$, is analogous
to a perturbation of the Schwarzschild metric studied by Moeckel\cite{moeckel}.
He described how a small body orbiting a Schwarzschild black hole causes
the inner orbits to become chaotic.

\
\begin{figure}[h]
\vspace{60mm}
\includegraphics{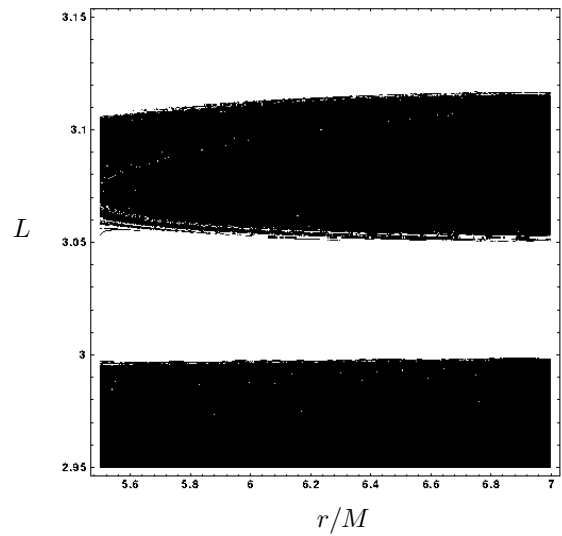}
\vspace{15mm}
\caption{Basin boundaries in the $(L,r)$ plane.}
\end{figure} 

\begin{picture}(0,0)
\put(5,150){$L$}
\put(120,40){$r/M$}
\end{picture}

Since our perturbed metric is static, the satellite's energy remains
a conserved quantity. In addition, the modified potential is
independent of $\theta$, so $p_\theta$ is conserved. This means we can
continue to study trajectories in the plane $\theta=\pi/2$ without
loss of generality. However, the $\phi$ dependence of the potential
breaks rotational invariance so the satellite's angular momentum is no
longer conserved. A trajectory with $L(0)>L_c$ can now evolve to one
with $L(\tau)<L_c$, thus allowing the possibility of capture by the black
hole. Conversely, a trajectory that was destined for captured might now
gain enough angular momentum to avoid capture.
For chaotic trajectories this gain or loss of angular momentum can
depend sensitively on initial conditions, leading to complicated
fractal boundaries separating the stable and unstable outcomes.

Here we are working under the usual assumption that we can use the
dynamics of the dissipationless system to predict which orbits will be
unstable when gravitational radiation is included as a form of
dissipation. This amounts to a kind of adiabatic approximation which
is valid when the energy loss per orbit is much less than the energy
of the satellite.

In Fig.~2 we display the stability basin in the $(L,r)$ plane for
a particle with energy $E=0.94$. A small pea with mass $m=M/100$ has
been introduced at $r=R=10M$. Particles are started from an
initial position $r$, with ``angular momentum'' $L=r^2U^2 d\phi /d\tau$.
The trajectory is then evolved numerically. If the particle is captured
by the black hole we colour the initial position white. If the particle
achieves a stable orbit (defined here to be 100+ orbits without being
captured), then we colour
the initial position black. There are two major features of note in
Fig.~2. The first is the irregular boundaries
between stable and unstable outcomes. The second is the broad swath of
unstable trajectories between $L=3M$ and $L=3.15M$. Before the pea was
introduced, all trajectories with $L>L_c=2.982M$ were stable. 

\
\begin{figure}[h]
\vspace{60mm}
 \includegraphics{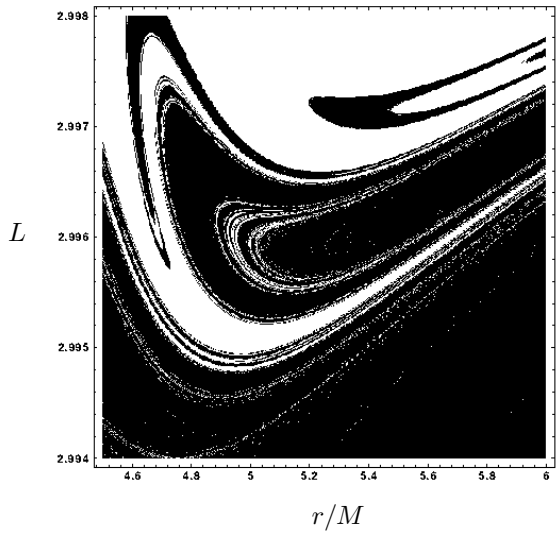}
\vspace{15mm}
\caption{A detail of Fig.~2 near $L=L_c$.}
\end{figure} 

\begin{picture}(0,0)
\put(5,150){$L$}
\put(120,43){$r/M$}
\end{picture}

The stability basins for the unperturbed spacetime are separated by a
smooth line at $L=L_c$. In Fig.~3 we display a detail of Fig.~2 which
clearly shows that the smooth boundary near $L_c$ has been replaced by a
complicated fractal structure. Note that most orbits with $L<2.995M$ have
been rendered unstable and thrown into the black hole. In Fig.~4 we
see that the central basin of instability also has a fractal boundary.
Fractal structures provide a gauge invariant signal of chaos in
general relativity\cite{us,sam,llama}.

But aside from producing nice pictures, what does chaos add to the
physics of satellites orbiting black holes? One interesting
possibility is an enhancement of the gravitational wave output.
Chaotic trajectories will tend to produce gravitational waves with
increased luminosity and amplitude, and erratic variations in amplitude.
The power radiated will exceed that of stable orbits since, in loose terms,
the luminosity is proportional
to the rate of change of acceleration. Chaotic orbits often exhibit
highly variable accelerations while nearly circular orbits have
gently varying acceleration profiles. In fact, even elliptical
orbits produce considerably more gravitational radiation than circular
orbits due to the sharp turn at their point of closest approach. Unfortunately,
most orbits are expected to be nearly circular by the time they
reach their swan song in the LIGO detection band\cite{ligo}.
This is because tidal
friction and gravitational radiation act to circularise the orbit.
To get more powerful gravitational wave signals
we need something to destabilise this picture. Enter chaotic resonances.

\
\begin{figure}[h]
\vspace{60mm}
\includegraphics{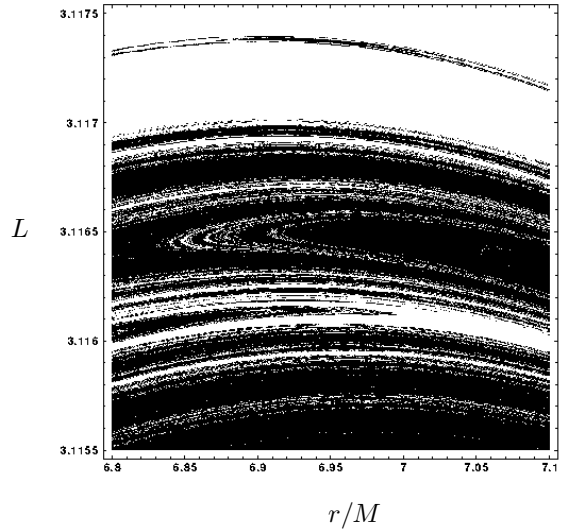}
\vspace{15mm}
\caption{A detail of Fig.~2 showing the boundary of the band of instability.}
\end{figure} 

\begin{picture}(0,0)
\put(5,157){$L$}
\put(125,49){$r/M$}
\end{picture}

As a realistic binary system spirals inwards, various relativistic
instabilities start to become important. The instabilities might be
caused by spin-orbit or spin-spin coupling, or by external mass distributions
such as the third body studied here. While these resonances are typically
restricted to isolated bands in phase space, the inspiralling satellite
is likely to run across an unstable band as its energy
and angular momentum are reduced by the emission of gravitational waves.
The resulting chaotic orbit would then provide a boost to the
gravitational wave output.

To get a feel for this chaotic enhancement we can calculate the wave
amplitude and power radiated using the quadrupole approximation.
Since the orbits we are studying are fairly relativistic,
the quadrupole approximation can only provide a qualitative picture.
The power radiated by the satellite is given by\cite{ll}
\begin{equation}
P={1 \over 5}<\! (d^3_t I_{ij})^2\!> \, ,
\end{equation}
where $<>$ denotes the average over several orbits and $I_{ij}$ is the
reduced quadrupole moment of the satellite's
orbit $(i,j=1,2,3)$:
\begin{equation}
I_{ij}= (x_{i}x_{j}-\frac{1}{3}r^2)\mu \, .
\end{equation}
The notation $d^{3}_{t}$ is shorthand for $d^3 / dt^3$. The
direction-averaged wave amplitude is given by
\begin{equation}
A\propto  \sqrt{(d^2_t I_{ij})(d^2_t I^{ij})} \, .
\end{equation}
Switching to cartesian coordinates, and considering orbits in
the plane $z=0$ $(\theta=\pi/2)$ we find
\[ A \propto \left( 3(d^{2}_{t}(xy))^2+[d^{2}_{t}x^2
-d^{2}_{t}y^2]^2+(d^{2}_{t}x^2)(d^{2}_{t}y^2)\right)^{1/2}  .
\]
In Fig.~5 we display the amplitude of three representative
trajectories with $E=0.94$, normalised against the stable circular
orbit of the unperturbed system with $L=3.1518M$.

\
\begin{figure}[h]
\vspace{60mm}
\includegraphics{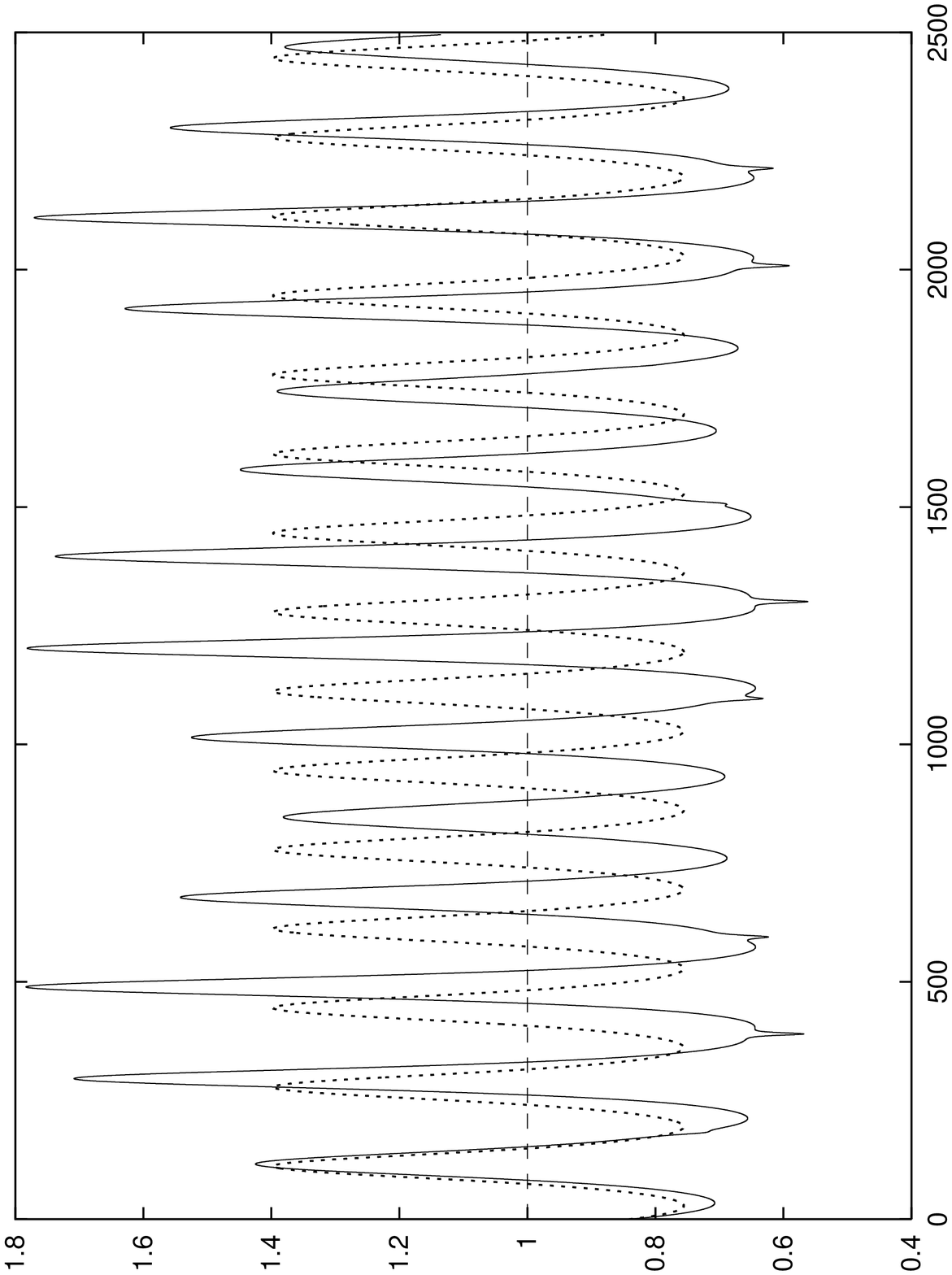}
\vspace{8mm}
\caption{The amplitude of circular (dashed line), elliptic (dotted curve)
and mildly chaotic (solid curve) satellite orbits.}
\end{figure} 

\begin{picture}(0,0)
\put(-8,220){$A$}
\put(120,60){$\tau$}
\end{picture}
\vspace*{-0.2in}

\noindent The dotted line shows the variation in amplitude
of a precessing elliptical orbit of the unperturbed system
with $L=3.117M$. The solid line is for an orbit with the same initial
conditions as the regular elliptic orbit, but this time for the
perturbed system. The variation in amplitude is increased by up to a
factor of two over the unperturbed system. The elliptic and chaotic
orbits have power outputs $1.48$ and $2.01$ times larger than the
stable circular orbit. This implies that the small perturbation caused
by the ``pea'' increased the power output by $36 \%$.

To better assess the significance of this result, we need to understand
the types of trajectories that give rise to increased power output,
how generic these might be in physically reasonable spacetimes and how
large the effect can be. To this end, we display in Fig.~6 a Poincar\'{e}
section for trajectories with $E=0.94$. The Poincar\'{e} section
reveals a combination of unbroken KAM tori, island chains, cantori and
thin stochastic layers. These features are typical for mildly chaotic
systems. The chaotic orbit shown in Fig.~5 forms the thin stochastic
layer that makes up the outermost ring of the Poincar\'{e}
section. While this is orbit is amongst the most chaotic we found,
it only ergodically wanders over a very small
band in phase space. The unbroken KAM tori prevent orbits from
becoming highly erratic. This in turn limits the gravitational wave
output. Moreover, orbits of the perturbed system that lie on the
unbroken KAM tori experience much smaller increases in power
output. For example, the outermost unbroken KAM tori in Fig.~6 produces
just $5\%$ more power than the analogous orbit of the unperturbed
spacetime.

\
\begin{figure}[h]
\vspace{60mm}
\includegraphics{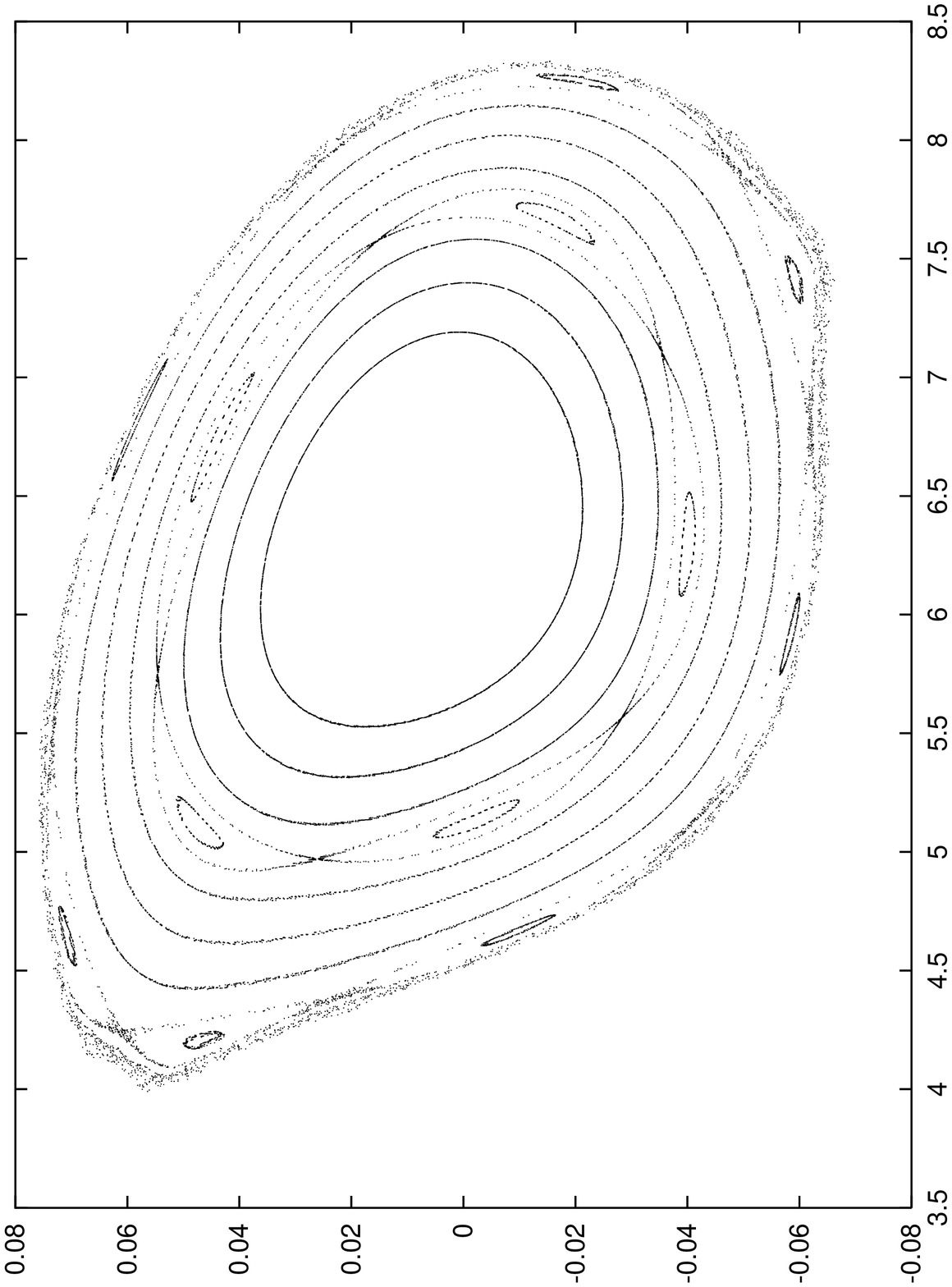}
\vspace{10mm}
\caption{A Poincar\'{e} section for trajectories with $E=0.94$.}
\end{figure} 

\begin{picture}(0,0)
\put(-10,150){$\partial_{\tau} r$}
\put(115,49){$r/M$}
\end{picture}
\vspace*{-0.2in}

These observations suggest that the power output is greatest for
orbits that fill the largest regions of phase space. In the model
spacetime we have been studying the dynamics is dominated by unbroken
KAM tori, and this limits the size of any stochastic layers. In
addition, we studied a 4-dimensional Hamiltonian system so the
unbroken KAM tori partition phase space. Systems with fewer KAM tori,
higher dimensional dynamics, and/or dissipation typically have trajectories
that wander over large regions of phase space. In such systems the
effects we have been describing will be larger and more widespread.

The literature already contains several realistic black hole models
where highly erratic orbits have been found. For example, some of the
trajectories shown in Fig.~1 of Ref.\cite{vl}, and Fig.~4(f) of
Ref.~\cite{maeda} form wide stochastic layers. We are currently
modelling these spacetimes to see how large the gravitational wave
enhancement can be.

Using a simple model we have shown that chaotic instabilities will
affect a satellite's transition from inspiral to plunge, and cause an
increase in gravitational wave production. Both of these features
could be important when producing gravitational wave templates for the
LIGO and LISA detectors.

We thank Sam Drake, Eric Poisson and Janna Levin for informative discussions.
This work was supported by the Australian Research Council.

\end{document}